\documentclass[final,twocolumn,3p,times]{elsarticle}

\usepackage{natbib}
\setcitestyle{square, comma, numbers,sort&compress, super}
\usepackage{caption}
\usepackage{graphicx}
\usepackage{subfig}
\usepackage{tabularx}
\usepackage{latexsym}
\usepackage{amstext,amsfonts,amssymb,amsmath}
\usepackage{bm}
\usepackage{todonotes}

\usepackage{jabbrv}

\newcounter{parcnt} 

\usepackage{nonfloat}
\usepackage{color}
\biboptions{comma,sort&compress}

\usepackage[compact]{titlesec}
\titlespacing{\section}{0pt}{1ex}{1ex}

\begin{document}
\begin{frontmatter}

\title{Extracting Information Overlap in Simultaneous OH-PLIF and PIV Fields with Neural Networks}

\author[fir]{Shivam Barwey\corref{cor1}}
\ead{sbarwey@umich.edu}
\author[fir]{Venkat Raman}
\author[sec]{Adam Steinberg}

\address[fir]{University of Michigan, 500 S State St, Ann Arbor 48109, USA}
\address[sec]{Georgia Institute of Technology, 620 Cherry Street, Atlanta 30332, USA}
\cortext[cor1]{Corresponding author:}

\begin{abstract}
Simultaneous measurements, such as the combination of particle image velocimetry (PIV) for velocity fields with planar laser induced fluorescence (PLIF) for species fields, are widely used in experimental turbulent combustion applications for the analysis of a plethora of complex physical processes. Such physical analyses are driven by the interpretation of spatial correlations between these fields by the experimenter. However, these correlations also imply some amount of intrinsic redundancy; the simultaneous fields contain overlapping information content. The goal of this work lies in the quantitative extraction of this overlapping information content in simultaneous field measurements. Specifically, the amount of PIV information contained in simultaneously measured OH-PLIF fields in the domain of a swirl-stabilized combustor is sought. This task is accomplished using machine learning techniques based on artificial neural networks designed to optimize PLIF-to-PIV mappings. It was found that most of the velocity information content could be retrieved when considering linear combinations of neighborhoods of OH-PLIF signal spanning roughly two integral lengthscales (half of the considered domain), and that PLIF signal interactions contained in smaller, local regions (less than half of the domain) contained no PIV information. Further, by visualizing the coherent structures contained within the neural network parameters, the role of multi-scale interactions related to velocity field retrieval from the OH-PLIF signal became more apparent. Overall, this study reveals a useful pathway (in the form of overlapping information content extraction) to develop diagnostic tools that capture more information using the same experimental resources by minimizing redundancy. 
\end{abstract}


\end{frontmatter}
\clearpage
\section{Introduction}
\label{sec:intro}

Laser diagnostics are powerful tools for studying complex physical processes in turbulent combustion, including turbulence/flame interactions, ignition, soot formation, flame lift-off, and thermoacoustic instabilities~\cite{Kohse2005,Barlow2007,Alden2011,Sick2013,Dreizler2015}. Many experiments utilize multiplexed diagnostics to simultaneously measure different quantities, such as the combination of particle image velocimetry (PIV) for velocity fields with planar laser induced fluorescence (PLIF) for species fields. This combination of species and velocity information leads to valuable physical insights regarding flow-flame interactions. 

The introduction of more than one type of measurement allows the practitioner to deduce the physical causes of instantaneous spatial (or spatio-temporal) correlations between these measurements. Deductions based on these correlations, however, are expert-guided, and a quantitative analysis of the information overlap between simultaneously measured fields due to this correlation is lost in the goal of understanding the underlying physics of the problem implied by the correlation itself. An important detail is that simultaneous fields that are highly correlated have overlapping information content, and thus contain redundancy. In particular, the information content of one type of measurement, say PIV, contained in another, say OH-PLIF, can in itself be valuable. 

The quantification of information overlap between simultaneously measured datasets is the focus of the current study. In particular, the velocity information contained in simultaneous OH-PLIF fields in the domain of a premixed swirl-stabilized combustor is obtained. The explicit quantification of this overlap can aid in developing new diagnostic combinations that produce simultaneous measurements that capture the same physical process, but minimize the overlapping content. Alternatively, it would allow the experimenter to design diagnostics that capture more information using the same experimental resources by minimizing redundancy. 

Techniques to obtain velocity fields from passive scalars relying on an inversion of the Navier-Stokes equations have been utilized in the past \cite{dahm1}. However, since OH is a reacting scalar, deducing velocity information directly from OH signals is more challenging and necessitates data-assisted approaches for information content retrieval. Such approaches revolve around the measure of statistical correlation between the simultaneous velocity and OH data \cite{adrian2000}. A standard way to compute correlations is to use image compression methods based on modal decompositions to extract important features of the flow. In the field of turbulent combustion, variants of proper orthogonal decomposition (POD) \cite{sirovich1987, steinbergPOD, duwig_ePOD}, dynamic mode decomposition (DMD) \cite{schmidDMD}, and spectral techniques \cite{palies2011} have been applied. In these approaches, the data is split into time or frequency-averaged spatial basis functions and time-varying coefficients. Standard implementations, however, are usually confined to either a single type of data or a concatenation of multiple types of simultaneously measured data. They do not take into account the ability to physically retrieve one field from another. 

While the approach used in this is related to modal decomposition and correlation, we also seek interpretations of these correlations. In this context, machine learning (ML) provides a useful framework. ML-based concepts have been used effectively in the context of turbulent combustion both recently \cite{ramanEmergingTrends,barwey_ctm,poinsot_cnn} and in previous decades \cite{masriANN, kempfANN, menonANN}. For example, we recently showed that an ML-based mapping function can be designed to take an input OH-PLIF field and transform it into a velocity field that accurately matches simultaneous three-component PIV data for complex reacting flows~\cite{shivam_cst}. The function was obtained using highly nonlinear deep convolutional neural networks. These ML techniques are widely known to be powerful, but their interpretation is often neglected in the pursuit of model accuracy.

The goal of this work is not to show that such a mapping can be achieved, but rather to identify {\it how/why} the mapping is achieved. For this purpose, two viewpoints are proposed:  a) a macroscopic viewpoint that seeks to explain the information overlap in terms of universality of turbulent structures and b) a microscopic viewpoint that relies on coherence between OH and velocity structures. It will be shown that a simplification of the regression model from Ref.~\cite{shivam_cst} into a simple artificial neural network (ANN) framework allows for a modal-decomposition based interpretation of the mapping function, similar to the methods described above. The key difference is that the decomposition is obtained by optimizing the PLIF-to-PIV mapping in a regression context, which allows for the simple, yet powerful ability to extract insight regarding the velocity information contained in the OH-PLIF field. 

\section{Dataset}
\label{sec:data}
\begin{figure}
    \centering
    \includegraphics[width=67mm]{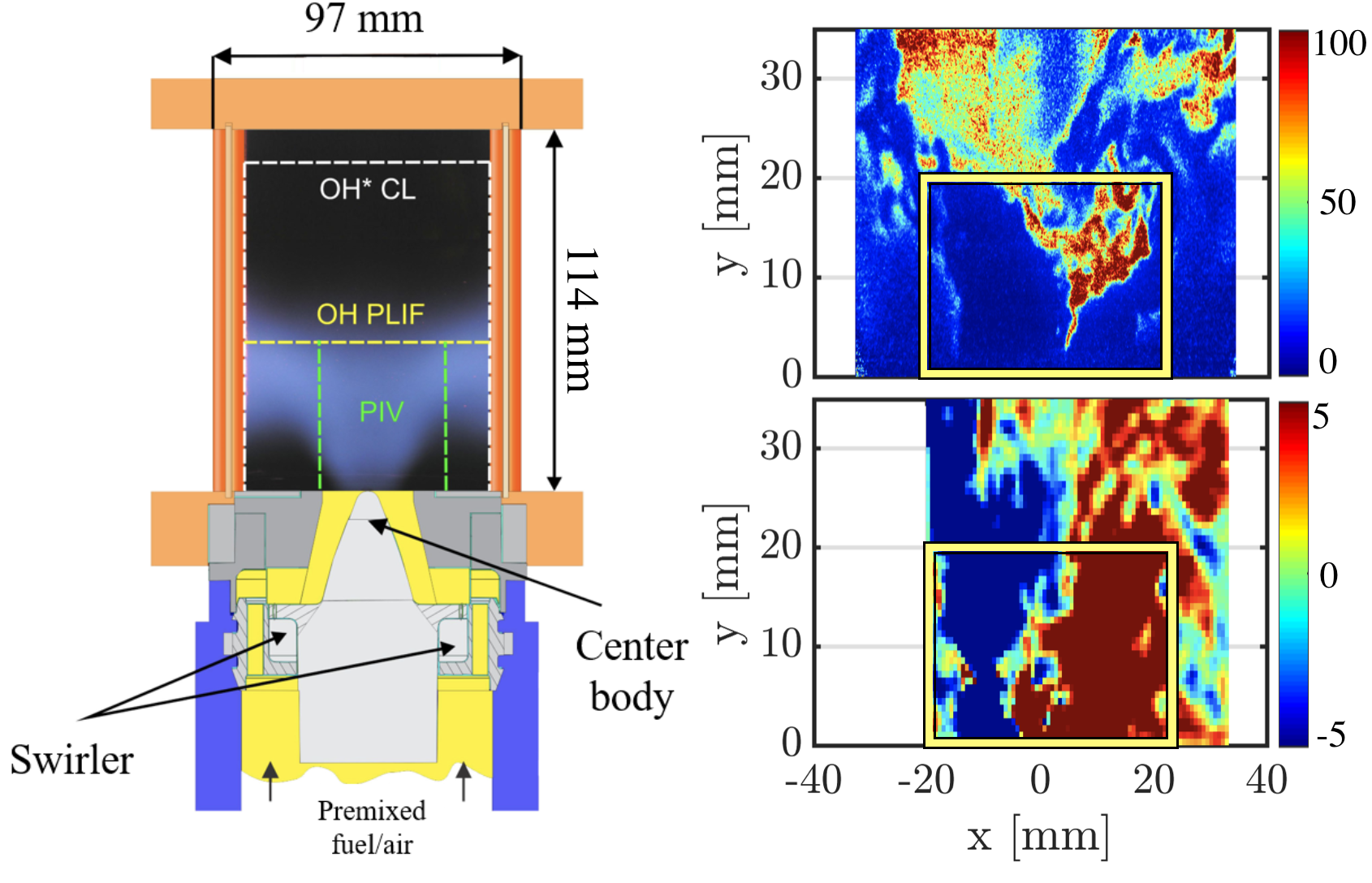}
    \caption{(Left) Combustor schematic. (Right) Samples of planar OH-PLIF (top) and PIV (bottom) images with cropped region indicated by box.}
    \label{fig:apparatus}
\end{figure}

Figure~\ref{fig:apparatus} shows a schematic of the gas turbine model combustor, which is similar to that originally described by Meier et al.~\cite{meier}, but with a slightly larger combustion chamber. Details of the experimental setup are given in An et al.~\cite{qiang,qiang_new}.

Premixed fuel and air are fed through a plenum to the radial swirler before entering the combustion chamber, where vortex breakdown generates a strong recirculation zone. The case studied here used a fuel consisting of 60\% CH$_4$ and 40\% CO$_2$ by volume, an equivalence ratio of $\phi = 0.60$, a preheat temperature of 400~K, and an air flow rate of 400~SLPM. At this operating condition, the combustor exhibits bimodal behavior, with the flame located either attached to the nozzle exit or lifted-off. Transition between the modes is spontaneous with no external changes to the operating conditions~\cite{qiang,qiang_new}. 

Data was collected using 10~kHz repetition-rate OH PLIF and stereoscopic PIV (S-PIV), providing time-resolved 2D measurements of the OH radical distribution and three-component velocity field. For this work, the dataset constitutes 3000 lifted flame snapshots: 2000 of these (a $0.2$~s time-series) are used for training and the remaining 1000 (a $0.1$~s time-series) for testing. Figure~\ref{fig:apparatus} shows a typical instantaneous OH-PLIF and PIV (out-of-plane component) snapshot in the lifted state. 

As a pre-processing step, for ease of cross-field analysis, the PLIF and PIV data were cropped as shown in Fig.~\ref{fig:apparatus} such that they 1) matched domain extent and 2) concentrated on the near-burner exit region, which captures much of the complexity of the lifted flame dynamics. This region was particularly interesting because there is significant velocity fluctuation for small amounts of OH-PLIF signal. As a last step, the finer-resolution OH-PLIF field was coarsened to match that of the PIV field, resulting in snapshot dimensions of $64\times 32$ pixels in the $x$- and $y$-directions respectively. 

\begin{figure}
    \centering
    \includegraphics[width=67mm]{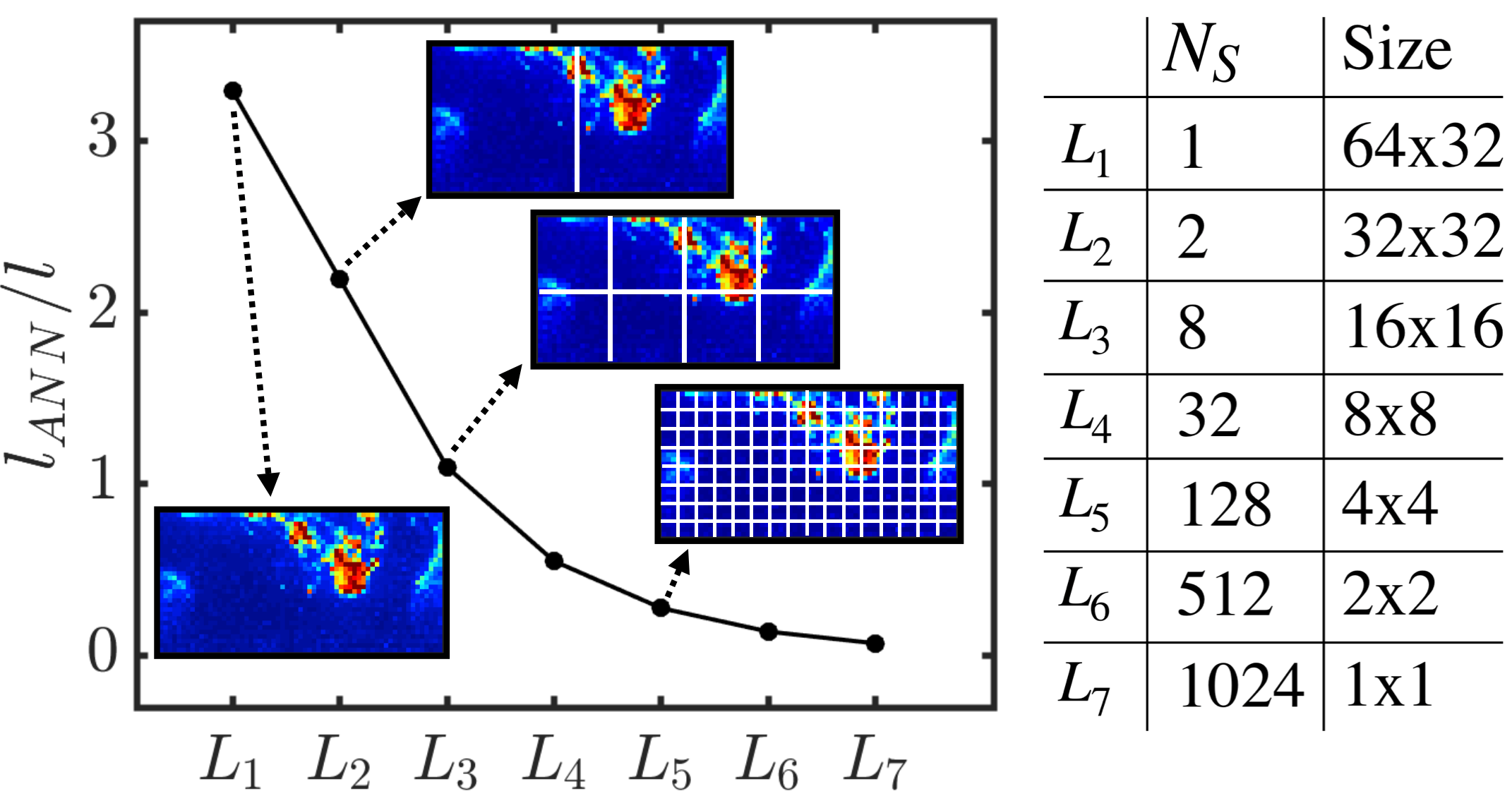}
    \caption{(Left) Subdomain lengthscale $l_{ANN}$ normalized by integral length scale $l$ for each $L_i$. (Right) Table listing number of subdomains per snapshot ($N_S$, first column) and subdomain size in pixels (second column) for each $L_i$.}
    \label{fig:lengthscales}
\end{figure}

In the macroscopic viewpoint, the relation between correlation of the PLIF-PIV images and turbulent length scales needs to be determined. To isolate lengthscale effects, each full-sized $64 \times 32$ snapshot was decomposed into a set of smaller non-overlapping subdomains, with the size of the domain providing a spatial length scale. This is useful for the following reasons. With an ANN model, the PIV output can be obtained in many different ways: one can choose to either a) provide the whole PLIF field as an input, in which case the network infers that a single PIV pixel prediction is a function of every single pixel in the PLIF input (which allows for the mapping function to capture non-local effects), or b) provide only a subsection of the PLIF input, in which case a given PIV pixel prediction is only a function of a subset of the collection of all PLIF pixels (the model is then implicitly constrained to local effects given by the size of the subsection). The mapping estimate corresponding to a particular lengthscale (or the neighborhood size of PLIF pixels chosen by the user to estimate the PIV field) then gives an overall indication of the required amount of spatial information content needed to achieve an accurate mapping. Seven different scales are used, denoted by $L_i, i=1, \ldots, 7$. Each $L_i$ is attributed a lengthscale based on its subdomain size, as shown in Fig.~\ref{fig:lengthscales}. The lengthscales shown in Fig.~\ref{fig:lengthscales} are normalized by the integral length scale, $l$, characterized by the shear layer width between the inflowing reactants and the central recirculation zone; $l=10$~mm for the conditions studied. 
 

\section{Methodology}  
\subsection{Artificial Neural Network Architecture}
\label{sec:ann}
The mapping goal is described by ${\bf y} = f({\bf x})$, where ${\bf x} \in \mathbb{R}^{M \times 1}$ is a (vectorized) PLIF input at some timestep, ${\bf y} \in \mathbb{R}^{M \times 1}$ is a simultaneously measured single-component PIV output, and $f$ is a stationary mapping function. The quantity $M$ represents both the input and output dimensionality, which is assumed to be the same for a given $L_i$. In this study, $f$ is obtained using traditional three-layer ANNs, which implies input/output relations of the following structure: 
\begin{equation}
\begin{split}
    \label{eq:ann}
     {\bf h} = \sigma(W_{PLIF}^T {\bf x}), \\ 
     {\bf \hat{y}} = W_{PIV} {\bf h}.
\end{split}
\end{equation}

\begin{figure}
    \centering
    \includegraphics[width=2.4in]{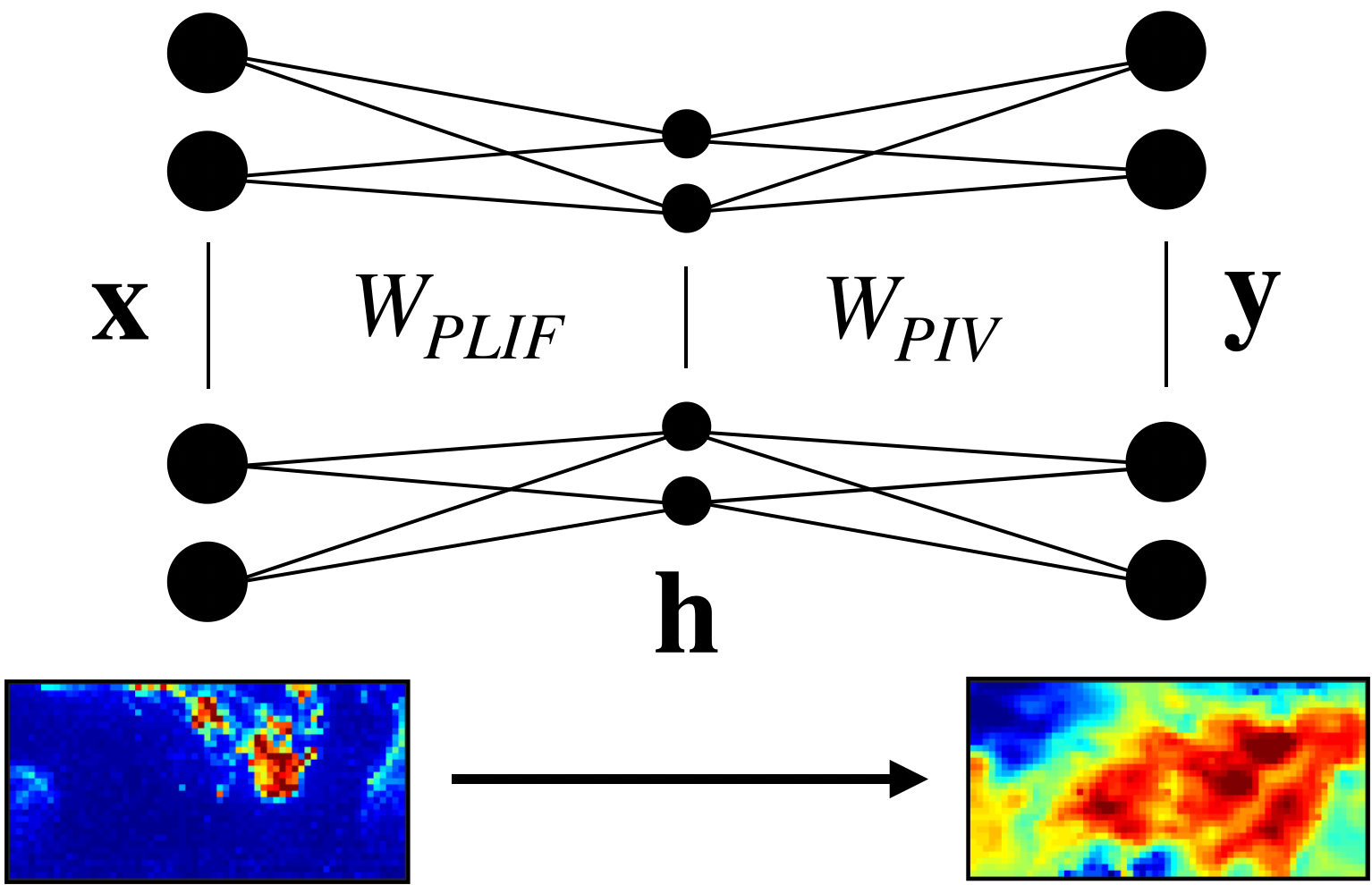}
    \caption{Illustration of three-layer ANN.}
    \label{fig:ann_schematic}
\end{figure}

Above, ${\bf h} \in \mathbb{R}^{N_H \times 1}$ is the hidden layer vector, $W_{PLIF},W_{PIV} \in \mathbb{R}^{M \times N_H}$ are the ANN weight matrices (parameters of the network), and ${\bf \hat{y}}$ is the predicted velocity field. The element-wise "activation" function $\sigma$ can be used to impose nonlinearity in the hidden layer representation of the input. If full linearity in the ANN desired, $\sigma$ is set to an identity map, which gives ${\bf \hat{y}} = W_{PIV} W_{PLIF}^T {\bf x}$. For the results below, only the linear model predictions are shown for two reasons. First, the nonlinear model provided nearly identical results. Second, the linear mapping has attractive properties for analysis that could be used to interpret the mapping. A schematic of the ANN is shown in Fig.~\ref{fig:ann_schematic}.

The weights are obtained using maximum likelihood estimation. In other words, $W_{PLIF},W_{PIV}$ are found such that Eq.~\ref{eq:ann} minimizes the mean-squared error over the predicted and exact PIV fields in the training set using gradient descent-based techniques.

In Sec.~\ref{sec:micro}, the data for each $L_i$ is fit with a unique ANN to extract macroscopic lengthscale effects relevant to the mapping. Since the quantity $M$ changes with $L_i$, a different ANN architecture must be used for each $L_i$. To be consistent with other modal decomposition techniques, the hidden layer dimension $N_H$ was set equal to rank of the training data matrix at each $L_i$. 

\subsection{Using the Cross-Covariance Matrix}
\label{sec:cross_covariance}
To understand how the mapping is performed at a level beyond general lengthscales, the ANN itself (i.e. the weights) must be interpreted. A pathway for interpretation is revealed if the ANN weights can be considered as a type of joint PLIF/PIV basis which can be visualized (columns of $W_{PLIF}$ and $W_{PIV}$ exist in the same phase space as the PLIF and PIV fields, and thus represent flow directions). This notion is made simpler by the linear ANN representation. However, a common practice is to randomly initialize the weights and leave them unconstrained during training -- there is no reason, then, to expect the converged weight vectors to be interpretable. 

In the three-layer linear model, the gradient descent dynamics of the weights during training are driven by the input/output cross-covariance matrix, $\Sigma_{YX}$. The singular value decomposition (SVD) of $\Sigma_{YX}$ then provides a good starting point for the ANN mapping interpretation. If the PLIF and PIV training data are concatenated by column in the matrices $X$ and $Y$ respectively, the cross-covariance matrix is $\Sigma_{YX} = YX^T$. Though deceptively simple in construction, it provides abundant information about cross-field mappings and has been used to great effect in geophysical applications \cite{svd_weather_book}. 

The SVD of $\Sigma_{YX}$ can be considered as the multi-field analog of the single-field POD, as it provides an orthogonal basis in which the covariance between the two fields, $X$ and $Y$, is diagonalized. The SVD is given by $\Sigma_{YX} = USV^T$. The coherent structures are contained in the $K$ basis vectors, or modes, which are the $M$-dimensional columns of $U \in \mathbb{R}^{M\times K}$ and $V\in \mathbb{R}^{M\times K}$. The modes corresponding to the high singular values in $S$ can be visualized to see which flow structures are responsible for a large majority of the data cross-covariance. The columns of $U$ and $V$ represent coherent structures in the PIV data (PIV modes) and PLIF data (PLIF modes), respectively. As such, the SVD of $\Sigma_{YX}$ enables multi-field feature extraction similar to the single-field counterpart enabled by POD \cite{parente_pca}.


Interestingly, it can be shown that the bases $U$ and $V$ assume an orthogonal PLIF-to-PIV mapping operation -- if $Y = AX$, then the solution to $A$ (in a least-squares sense) with an orthogonal restriction on the PLIF/PIV transformation is provided by $A = U V^T$ \cite{svd_weather_book}. Recall from Sec.~\ref{sec:ann} that in the linear ANN case, $A = W_{PIV} W_{PLIF}^T$. Thus, if $W_{PIV}$ and $W_{PLIF}$ are set equal to $U$ and $V$ (which assumes $N_H=K$), the $M$-dimensional singular vectors contained in the matrices $U$ and $V$ can be interpreted as types of constrained ANN \textit{weight vectors}, where the weight vectors are the $M$-dimensional columns of $W_{PLIF}$ and $W_{PIV}$. There is, however, no reason to expect the $A = U V^T$ mapping to be optimal, since it assumes that the PLIF/PIV weights which compose the linear mapping in $A$ are orthonormal; physical interpretation of the singular vectors of $\Sigma_{YX}$ in the context of complex field-to-field mappings should therefore be treated with great care. However, it will be shown that if the ANN weights $W_{PLIF}$ and $W_{PIV}$ are \textit{initialized} with $U$ and $V$ instead of random weight matrices, a more interpretable model can be recovered, and further insight to the mapping procedure can be obtained.

\section{Results} 
\label{sec:results}
Two viewpoints are used here to pinpoint the concentration of information content used in the PLIF to PIV field transformation. In the first, the role of spatial information content is assessed from a macroscopic viewpoint. The goal is to quantify the length scale at which cross-domain interactions play a role in the mapping procedure. To achieve this, different ANNs are trained for the different domain lengthscales discussed in Sec.~\ref{sec:data} and are analyzed. The second viewpoint is a microscopic viewpoint. Here, using the $L_1$ model, the goal is to investigate exactly how the ANN achieves the mapping by identifying specific regions within the PLIF domain most relevant to the predicted velocity field. Results from the second viewpoint will then be connected to the findings obtained from the first. 

\subsection{Macroscopic Viewpoint} 
\label{sec:macro}

Following Ref.~\cite{dahm1}, the standard moment correlations between the predicted and exact velocity fields are used here to assess model performance for the ANNs, given by 
\begin{equation}
    R_{{\bf y}, {\bf \hat{y}}} = \frac{\langle {\bf y}' {\bf \hat{y}}'  \rangle} {{\bf y}'_{RMS} {\bf \hat{y}}_{RMS}'}. 
\end{equation}



\begin{figure}
    \centering
    \includegraphics[width=0.45\textwidth]{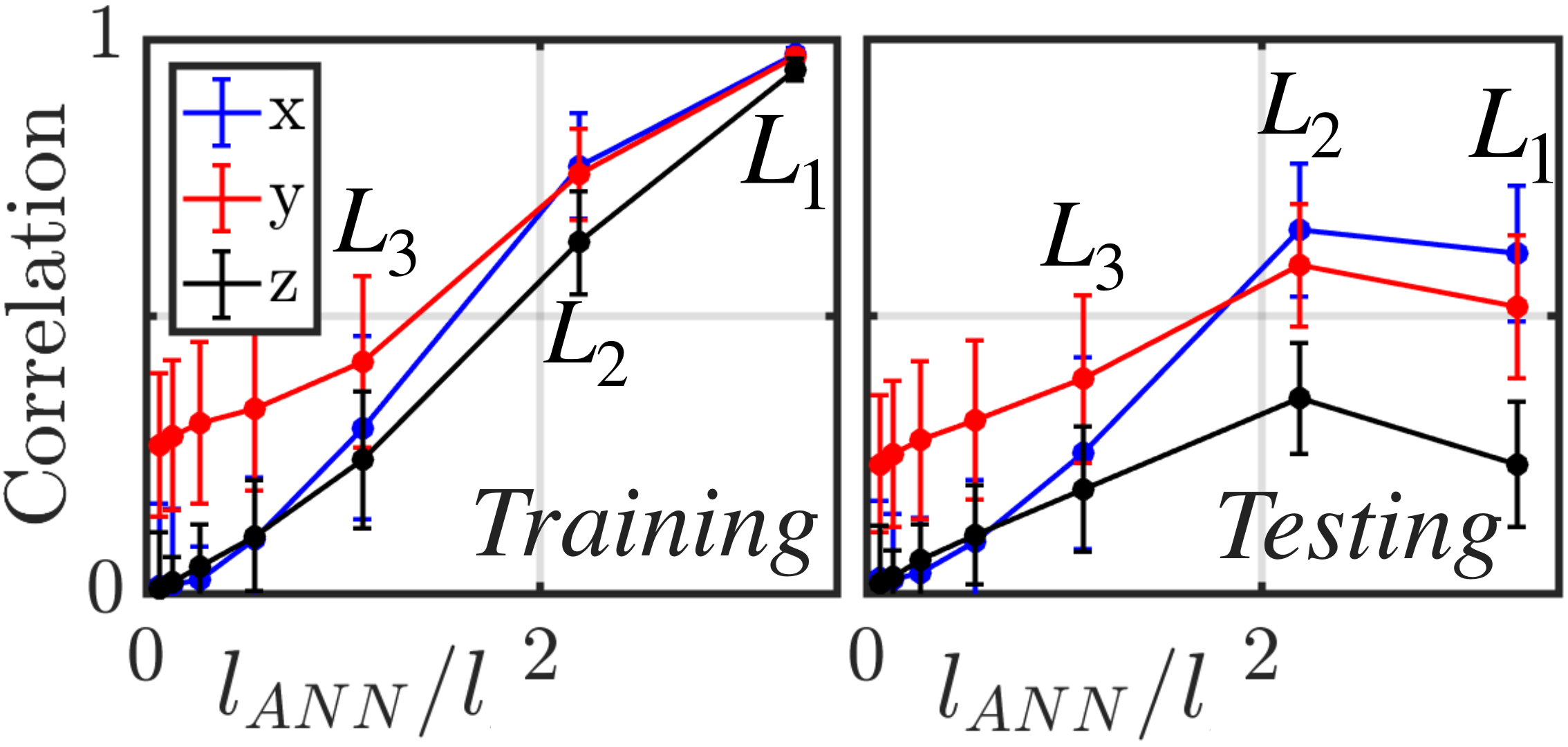}
    \caption{Mean correlation magnitudes between predicted/exact PIV fields as a function of ANN lengthscale. ANN lengthscale normalized by nozzle diameter, with highest three ANN scales shown on top for clarity.}
    \label{fig:ite_plot}
\end{figure}

Figure~\ref{fig:ite_plot} shows the average correlation magnitudes for all three components, at all seven tested lengthscales, for the training and testing sets. For the training set, all three velocity components display monotonically increasing correlation with respect to model lengthscale. For the testing set, correlations are overall lower and the trends are quite different. The testing set values peak in the average correlation at the $L_2$ model. Recall from Fig.~\ref{fig:lengthscales} that the $L_2$ model utilizes half of the PLIF domain as the input, with this input spanning roughly 2 integral lengthscales. However, since the standard deviations in the correlations are quite high, it can be concluded that both $L_1$ and $L_2$ models capture similar predictive ability for the testing set. Interestingly, the PIV-x and y models display significantly more improvement from higher levels of spatial information content than the PIV-z counterpart, and PIV-z performance is lowest across the board. This implies that non-local correlations play a more significant role in the PIV-x and y mappings than in the PIV-z mappings. This notion will be explored further in Sec.~\ref{sec:micro}. 

\begin{figure}
    \centering
    \includegraphics[width=0.45\textwidth]{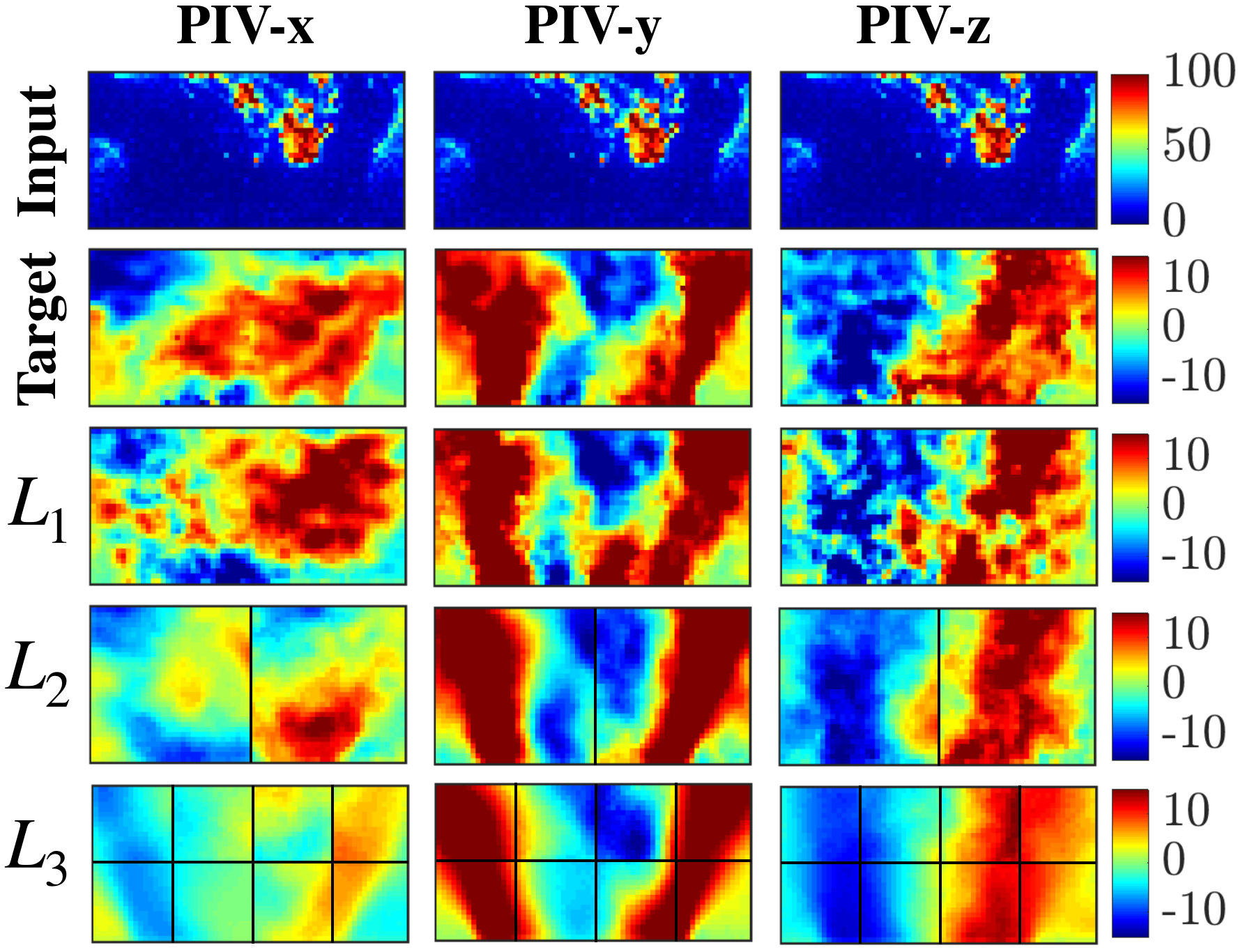}
    \caption{Examples of predicted velocity fields generated by a single OH-PLIF field (first row) from testing set using $L_1$, $L_2$, and $L_3$ ANNs. Columns correspond to PIV-x, y, and z predictions respectively. PIV units in m/s, PLIF in units of pixel intensity.}
    \label{fig:predictions}
\end{figure}

The spatial correlation between predicted and exact velocity fields in the testing set (values in Fig.~\ref{fig:ite_plot}) is not perfectly indicative of similarity in the structure of the instantaneous predicted fields themselves. Predicted instantaneous velocity fields for an example PLIF field input for the $L_1$, $L_2$, and $L_3$ ANN models is shown in Fig.~\ref{fig:predictions}. Despite seemingly low correlations in Fig.~\ref{fig:ite_plot} (especially in PIV-z), the predictions, though not perfect, appear reasonable. The $L_3$ model shows smoothed velocity streaks -- it essentially assigns the mean PIV field to all input PLIF fields and neglects fluctuations. The same is true for $L_4$ through $L_7$ (not shown), but to a much greater degree. The $L_2$ and $L_1$ model, however, show much more accurate solutions consistent with the trends observed in Fig.~\ref{fig:ite_plot}. Despite the fact that the $L_1$ model appears to be more physically accurate, it is motivating to see that reasonable velocity field predictions can be obtained with the $L_2$ model that utilizes only half of the available OH-PLIF signal. 

Ultimately, results from Figs.~\ref{fig:ite_plot} and \ref{fig:predictions} show how highly localized models that utilize OH content which spans length scales smaller than the turbulence integral scale (i.e. models $L_3$ onwards) do not contain the PIV fluctuation information. In other words, a potential lower-bound on the size of PLIF features most relevant to the mapping is in the $L_2$ model, which corresponds to half of the input PLIF field and captures 2 integral lengthscales worth of spatial content. This essentially points to the role of non-local, domain-wide PIV information contained in the PLIF field. From this, one can surmise either a) the structures which contribute to most of the mapping accuracy are large-scale in nature, b) the small-scale interactions in the domain play a role, but these interactions are non-localized, or c) a combination of these. 


\subsection{Microscopic Viewpoint} 
\label{sec:micro}
In order to gain a better understanding of the ANN mapping and confirm the aforementioned postulations in the end of Sec.~\ref{sec:macro}, a more detailed inspection of the model is warranted. Recall from Sec.~\ref{sec:cross_covariance} the notion of initializing weights with singular vectors of the cross-covariance before training. First, the advantages of using this informed initialization (also known as \textit{pre-training}) will be discussed and explicitly shown. Then, the weights themselves will be analyzed. The $L_1$ model is exclusively used here, as visualization of information content in the whole domain is sought. 

\begin{figure}
    \centering
    \includegraphics[width=67mm]{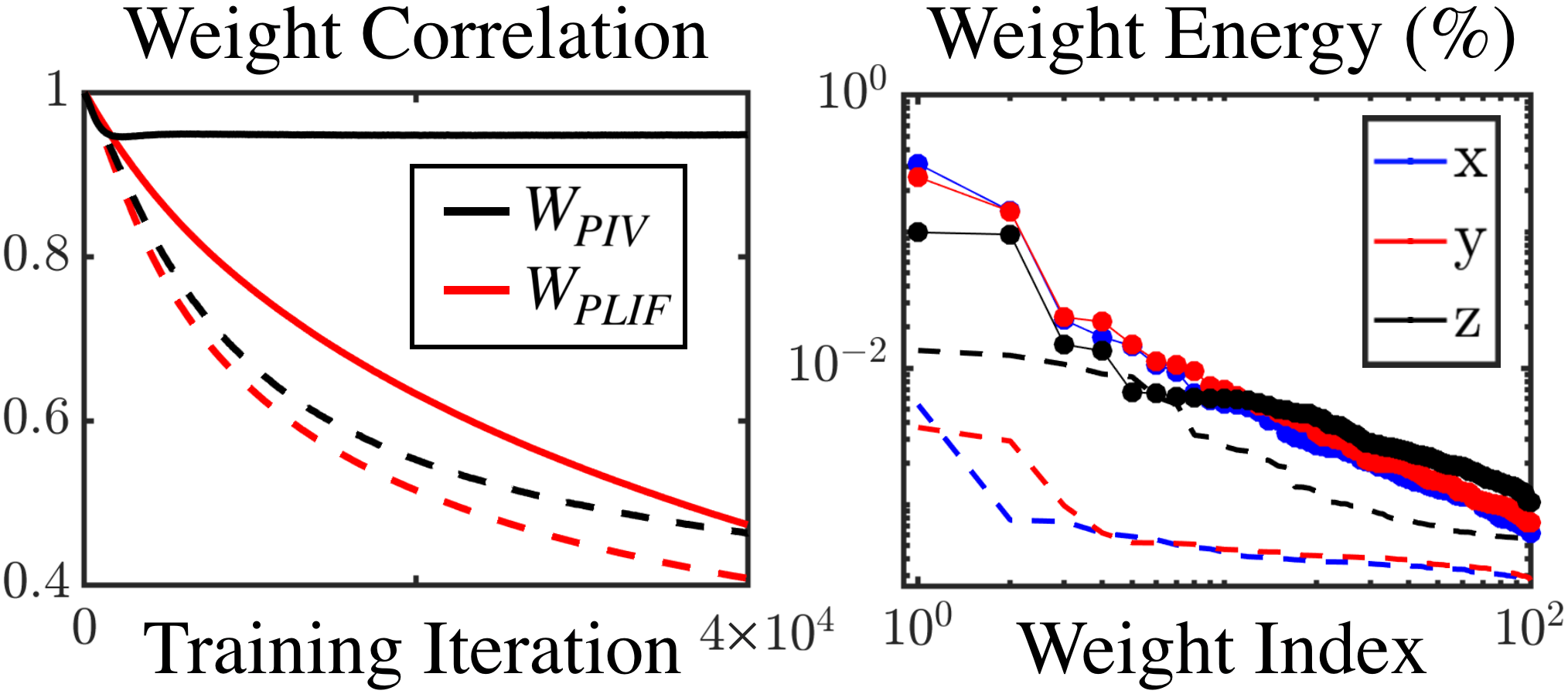}
    \caption{(Left) Average correlation of weight vectors with initial conditions versus training iteration. (Right) Variance of hidden layer contained in the input (first 100 weight indices shown sorted by descending energy). In both plots, dotted line is random initialization, solid is SVD-based initialization.}
    \label{fig:weight_energy}
\end{figure}

Assuming the loss function decreases during training (which indicates that the mapping implied by the initial set of weights is non-optimal), one can attribute any deviation in the weights from the initial condition during training to overall improvement in mapping accuracy. This tracking of weight deviation from the initial conditions is shown in the left plot of Fig.~\ref{fig:weight_energy} for the PIV-x model (same trends for others). Plotted on the y-axis is the average correlation between the weights and their initial condition as a function of training iteration, for both randomly initialized (dotted line) and SVD-initialized (solid line) neural networks. Note that, though not shown here due to space restriction, both the random and SVD-initialized models produced near-identical mapping results. 

For the randomly initialized weights, Fig.~\ref{fig:weight_energy} shows how both PLIF and PIV weights are largely changed from the initial state, implying that the mapping accuracy is distributed evenly throughout all parameters, which renders interpretation difficult. For the SVD-initialized weights, much of the change associated with the decreasing loss function (i.e. mapping accuracy) is due in large part to modification of $W_{PLIF}$ only. The $W_{PIV}$ values are largely unchanged, which is useful interpretation as will be seen below.

The right plot of Fig.~\ref{fig:weight_energy} shows the distribution of variance in the PLIF data contained in the hidden layer, which is a type of importance ranking for the PLIF weight vectors. The figure shows how initialization can be interpreted as a re-distribution of the converged hidden layer variance: in contrast to the randomly initialized ANN, most of the hidden layer variation is captured in only the first two weight vectors of $W_{PLIF}$ for the SVD-initialized ANN in all three velocity components. This helps with interpretability, since only a few of the weight vectors in the ANN are needed to characterize model behavior. 

\begin{figure}
    \centering
    \includegraphics[width=50mm]{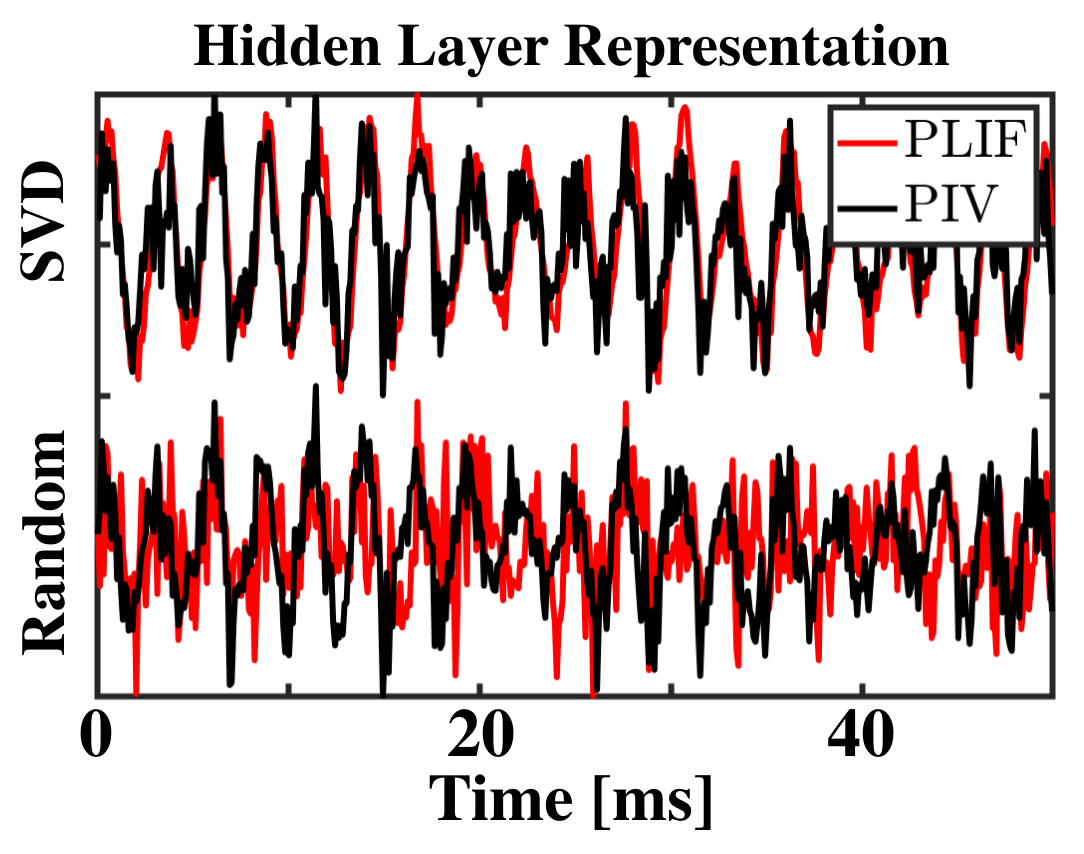}
    \caption{Testing set representation of the highest energy hidden layer node for models using SVD of $\Sigma_{XY}$ (top) and randomly (bottom) initialized weights.}
    \label{fig:hidden_values}
\end{figure}

With this in mind, the results from Fig.~\ref{fig:weight_energy} ultimately imply that 1) starting from the initial condition of the singular vectors of $\Sigma_{YX}$, only the singular vectors associated with the PLIF field were significantly modified, and 2) the fact that the high-energy singular vectors of the PIV field were \textit{not} largely modified implies that the ANN preserves much of the orthogonality of the $W_{PIV}$ initial condition. This second point is crucial. It allows for the high-energy hidden layer values to be recovered by $W_{PIV}$ as well as $W_{PLIF}$, even in a feed-forward architecture, via the transpose $W_{PIV}^T$. This agreement is shown for the first component of the hidden layer vector in Fig.~\ref{fig:hidden_values}. The red lines in Fig.~\ref{fig:hidden_values} show the true hidden layer values generated by the input due to $W_{PLIF}$, and the black lines show the same hidden layer values generated by the output PIV field using $W_{PIV}^T$. The agreement in PLIF/PIV hidden layer representation for the SVD-initialized models means that the variance distribution of the hidden layer due to the PLIF input (right plot of Fig.~\ref{fig:weight_energy} can also be attributed to the PIV output for the high-energy weight vectors. As a result, the first few weight vectors (columns) in $W_{PLIF}$ and $W_{PIV}$ are intrinsically connected, and can therefore be visualized to access the spatial distribution of PIV information content in the PLIF domain. 

\begin{figure}
    \centering
    \includegraphics[width=67mm]{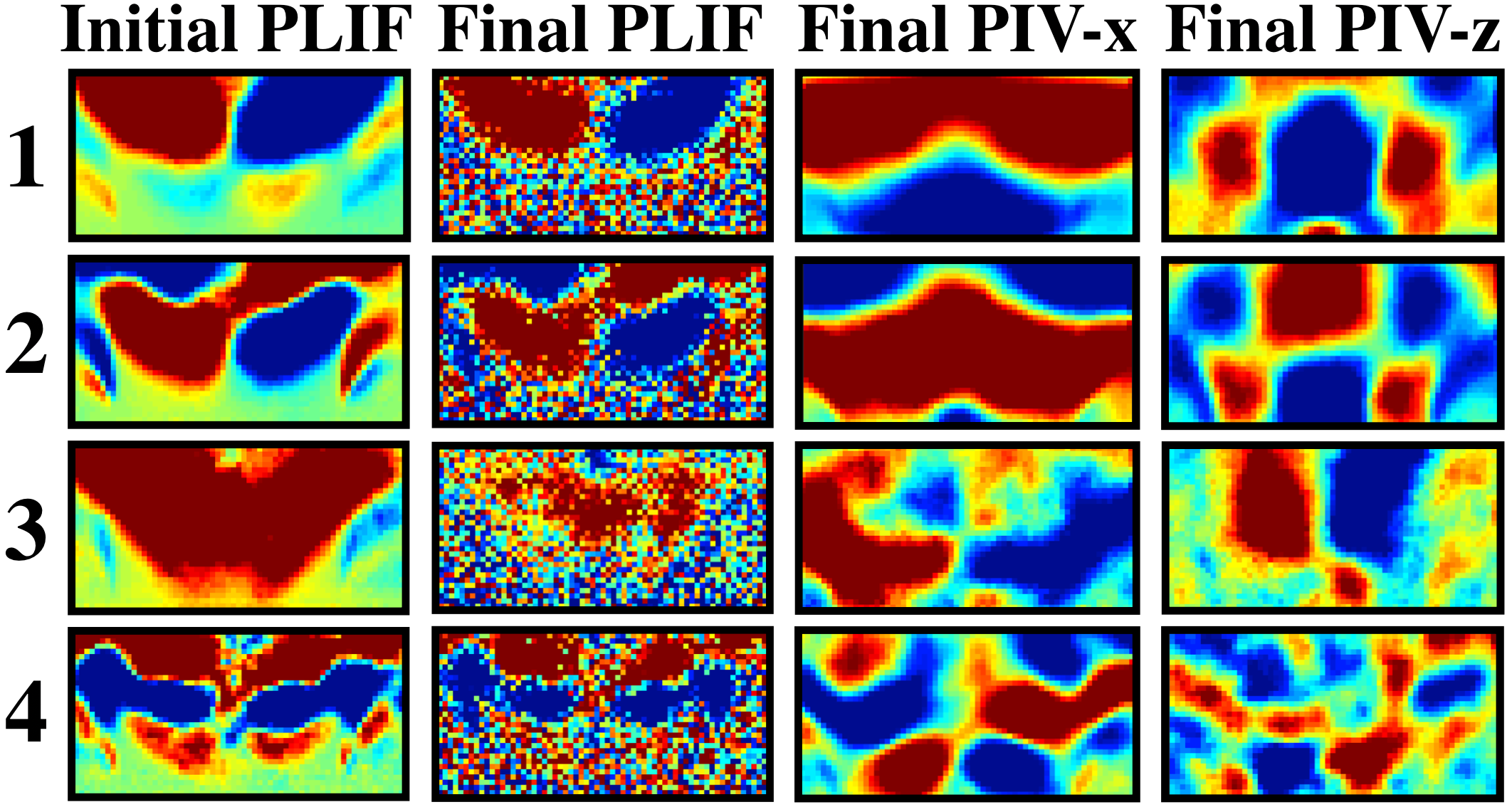}
    \caption{The four highest-energy ANN weight vectors as per Fig.~\ref{fig:weight_energy} for the SVD-initialized models. Rows indicate weight index. All colormaps scaled evenly; positive is red and negative is blue.}
    \label{fig:ann_modes}
\end{figure}

The four highest-energy ANN weights from Fig.~\ref{fig:weight_energy} are shown in Fig.~\ref{fig:ann_modes}. Only the final PLIF weights are compared to the corresponding initial conditions, since the final PIV weights were nearly identical to their initial conditions. Further, due to similarities in trends, PIV-y model weights are not shown. 

Note that the first two weight vectors (first two rows in Fig.~\ref{fig:ann_modes}) are most significant as per Fig.~\ref{fig:weight_energy}. Interestingly, all three models (PIV-x, y, and z) produced nearly identical coherent structures in the high-energy PLIF weights, implying that most of the velocity information content stored in the PLIF field is independent of velocity component. For all models, the large-scale structures in the final PLIF weights in Fig.~\ref{fig:ann_modes} appear to concentrate on regions where the OH signal is present in the detached flame dynamics, which may seem obvious. What is less obvious is that the changes required to achieve optimal mapping accuracy stems from the addition of small-scale artifacts in the lower half of the PLIF domain -- sparsity is completely removed in the final PLIF weights. The PLIF weights show how the velocity information content is stored in the interactions between the large coherent structures in the upper domain and the small incoherent fluctuations in the lower domain, implying clear multi-scale behavior. 

Additionally, there is evident symmetric structure in the ANN weights. For example, despite the asymmetry of the actual velocity fields, the dominant PIV weights (1 and 2) recovered from the ANN are symmetric and the PLIF weights are anti-symmetric (in the large structures). This explains why Fig.~\ref{fig:ite_plot} revealed significant spikes in the $L_2$ model, as only half of the domain is required to reveal the large-scale information content structure relevant for the mapping. This is consistent with the dynamics of the flame anchoring point imposed by the precessing vortex core, which is periodic about the x=0 centerline. The symmetric/anti-symmetric structure appears to falter earlier for higher weight numbers in PIV-z, which may be an indicator for its lower correlations observed in Fig.~\ref{fig:weight_energy}. On a similar note, the high PIV-x accuracies from Fig.~\ref{fig:ite_plot} may be related to the fact that, for the same high-energy PLIF weights (1 and 2), the corresponding PIV weight is more "excited". For example, the same excited regions in PLIF weight 1 corresponds to more excitation in PIV-x weight 1 than for PIV-z weight 1. The weight vectors thus show both the structures of velocity information content in the PLIF field through $W_{PLIF}$, as well as the impact felt by these structures on the velocity field through the coherent features in $W_{PIV}$. 

\section{Conclusion} 
Attributes of the PIV information content contained in the simultaneously measured PLIF field were extracted from the perspective of two viewpoints. The \textit{macroscopic} viewpoint, which considered an ensemble of ANNs trained for unique lengthscales, was used to isolate the degree at which local spatial PLIF information can be used to produce the PIV field. It was found that accurate velocity decodings required input PLIF content to span at least two integral lengthscales. This implies that most of the velocity information is contained in roughly twice the integral lengthscale, and that PLIF signal interactions required to obtain an accurate mapping are domain-wide in nature. To build on this, the \textit{microscopic} viewpoint allowed for a closer inspection of the spatial distribution of PIV information content by creating interpretable ANN weights. By tracking the change in these weights from their initial conditions, it was found that the velocity information was captured by adding small-scale OH-PLIF interactions in the lower domain to pre-existing large-scale coherent structures in the upper domain, supplementing the findings from the macroscopic analysis. 

The study shows how simple neural network models can be used to isolate redundancy in simultaneous measurements via the tracking of information content. The extension of this analysis to nonlinear deep neural network will be considered in future work. Ultimately, this study reveals a pathway for experimentalists to develop diagnostic tools that capture more information using the same experimental resources by minimizing redundancy. 

\bibliographystyle{jabbrv_unsrt}
\bibliography{sample}

\begin{thebibliography}{10}

\bibitem{Kohse2005}
Katharina Kohse-H{\"o}inghaus, Robert~S Barlow, Marcus Ald{\'e}n, and
  J{\"u}rgen Wolfrum, {\em\JournalTitle{Proceedings of the Combustion
  Institute}}, 30(1):89--123,  2005.

\bibitem{Barlow2007}
Robert~S Barlow, {\em\JournalTitle{Proceedings of the Combustion Institute}},
  31(1):49--75,  2007.

\bibitem{Alden2011}
Marcus Ald{\'e}n, Joakim Bood, Zhongshan Li, and Mattias Richter,
  {\em\JournalTitle{Proceedings of the Combustion Institute}}, 33(1):69--97,
  2011.

\bibitem{Sick2013}
Volker Sick, {\em\JournalTitle{Proceedings of the Combustion Institute}},
  34(2):3509--3530,  2013.

\bibitem{Dreizler2015}
A~Dreizler and B~B{\"o}hm, {\em\JournalTitle{Proceedings of the Combustion
  Institute}}, 35(1):37--64,  2015.

\bibitem{dahm1}
Lester~K Su and Werner~JA Dahm, {\em\JournalTitle{Physics of Fluids}},
  8(7):1869--1882,  1996.

\bibitem{adrian2000}
MG~Olsen and RJ~Adrian, {\em\JournalTitle{Experiments in fluids}},
  29(1):S166--S174,  2000.

\bibitem{sirovich1987}
Lawrence Sirovich, {\em\JournalTitle{Quarterly of Applied Mathematics}},
  45(3):561--571,  1987.

\bibitem{steinbergPOD}
Adam~M Steinberg, Isaac Boxx, Michael St{\"o}hr, Campbell Carter, and Wolfgang
  Meier, {\em\JournalTitle{Combustion and Flame}}, 157(12):2250--2266,  2010.

\bibitem{duwig_ePOD}
Christophe Duwig and Piero Iudiciani, {\em\JournalTitle{Flow, turbulence and
  combustion}}, 84(1):25,  2010.

\bibitem{schmidDMD}
Peter~J Schmid, {\em\JournalTitle{Journal of Fluid Mechanics}}, 656:5--28,
  2010.

\bibitem{palies2011}
Paul Palies, Daniel Durox, Thierry Schuller, and S{\'e}bastien Candel,
  {\em\JournalTitle{Combustion and Flame}}, 158(10):1980--1991,  2011.

\bibitem{ramanEmergingTrends}
Venkat Raman and Malik Hassanaly, {\em\JournalTitle{Proceedings of the
  Combustion Institute}}, 37,  2019.

\bibitem{barwey_ctm}
Shivam Barwey, Malik Hassanaly, Qiang An, Venkat Raman, and Adam Steinberg,
  {\em\JournalTitle{Combustion Theory and Modelling}}, pages 1--27,  2019.

\bibitem{poinsot_cnn}
CJ~Lapeyre, A~Misdariis, N~Cazard, D~Veynante, and T~Poinsot,
  {\em\JournalTitle{Combustion and Flame}},  Submitted.

\bibitem{masriANN}
FC~Christo, AR~Masri, and EM~Nebot, {\em\JournalTitle{Combustion and Flame}},
  106(4):406--427,  1996.

\bibitem{kempfANN}
A~Kempf, F~Flemming, and J~Janicka, {\em\JournalTitle{Proceedings of the
  Combustion Institute}}, 30(1):557--565,  2005.

\bibitem{menonANN}
Baris~Ali Sen and Suresh Menon, {\em\JournalTitle{Combustion and Flame}},
  157(1):62--74,  2010.

\bibitem{shivam_cst}
Shivam Barwey, Malik Hassanaly, Venkat Raman, and Adam Steinberg,
  {\em\JournalTitle{Combustion Science and Technology}}, pages 1--24,  2019.

\bibitem{meier}
WEIGAND Meier, P~Weigand, XR~Duan, and R~Giezendanner-Thoben,
  {\em\JournalTitle{Combustion and Flame}}, 150(1-2):2--26,  2007.

\bibitem{qiang}
Qiang An, Wing~Yin Kwong, Benjamin~D Geraedts, and Adam~M Steinberg,
  {\em\JournalTitle{Combustion and Flame}}, 168:228--239,  2016.

\bibitem{qiang_new}
Qiang An and Adam~M Steinberg, {\em\JournalTitle{Combustion and Flame}},
  199:267--278,  2019.

\bibitem{svd_weather_book}
Antonio Navarra and Valeria Simoncini,  {\em A guide to empirical orthogonal
  functions for climate data analysis}.

\bibitem{parente_pca}
James~C Sutherland and Alessandro Parente, {\em\JournalTitle{Proceedings of the
  Combustion Institute}}, 32(1):1563--1570,  2009.

\end{thebibliography}
\end{document}